\documentclass[prl,showpacs,twocolumn,aps]{revtex4-1}

\usepackage{latexsym}
\usepackage{amssymb}
\usepackage{amsfonts}
\usepackage{epsfig}
\usepackage{psfrag}
\usepackage{graphicx}

\usepackage{epsfig,color}
\usepackage{psfrag}
\usepackage[utf8]{inputenc} 
\definecolor{darkgreen}{rgb}{0,0.4,0}

\newcommand{\bea}{\begin{eqnarray}}
\newcommand{\ea}{\end{eqnarray}}
\newcommand{\eea}{\end{eqnarray}}

\usepackage{amsmath}

\usepackage{color}
\renewcommand{\ll}{\langle\langle}

\begin{document}

\title{Entangling continuous variables with a qubit array}


\author{Patrick Navez$^{1,2}$, Artur Sowa$^{2}$, 
Alexander Zagoskin$^{3}$}

\affiliation{$^1$ 
Helmholtz-Zentrum Dresden-Rossendorf, Bautzner Landstra{\ss}e 400,
01328 Dresden, Germany 
\\
$^2$ 
University of Saskatchewan, Dept of Math. and Stat. , Saskatoon, S7N 5E6, Canada
\\
$^3$
Dept of Phys., Loughborough University,
Loughborough LE11 3TU, United Kingdom
}

\date{\today}

\begin{abstract}
We show that an array of qubits embedded in a waveguide can emit entangled pairs of microwave photon beams. The quadratures obtained from a  homodyne detection of  these outputs beams form a pair of  correlated continuous variables similarly to the EPR experiment. The photon pairs are produced by the decay of plasmon-like collective excitations in the qubit array. The maximum intensity of the resulting beams is only bounded by the number of emitters. We calculate the  excitation decay rate both into a continuum of photon state  and into a one-mode cavity. We also determine the frequency of Rabi-like oscillations resulting from a detuning. 
\end{abstract}


\maketitle
{\it Introduction:}
The steady improvement of superconducting electronics over the last two decades \cite{PhysRevLett.107.240501,PhysRevA.76.042319,wendin2017quantum} cemented the place of Josephson effect-based devices among the leading platforms for quantum technologies (e.g., quantum computation) \cite{RevModPhys.73.357}. However, the control and observation of essential quantum correlations and entanglement necessary for the operation of these technologies remains a challenging task \cite{zagoskin2014test,PhysRevB.95.064304}. A convenient testbed for this research is provided by a superconducting qubit array embedded in a coplanar waveguide (a "1D quantum metamaterial" setup) \cite{PhysRevLett.107.240501,PhysRevA.76.042319,wendin2017quantum,wallraff2004strong,PhysRevLett.91.097906,Hoi_2013,artur}. Some of  these metamaterials are predicted to display interesting nonlinear properties like the two-photon induced transparency \cite{rakhmanov2008quantum,zagoskin2009quantum,savel2012two}, superradiance 
\cite{bamba2016superradiant,asai2018quasi} and lasing \cite{ivic2016qubit} but these results have been obtained using approximations short of a full QED treatment.
  
In this Letter we build a consistent theory for a linear array of qubits placed in a waveguide. Specifically, we consider a set of capacitively coupled transmons, but the general results are not going to be sensitive to the particular kind of a qubit. The Josephson junctions are arranged symmetrically in order to ensure a quadratic coupling to the electromagnetic field. The collective excitations are produced by abrupt changes of qubit electric charges, tantamount to a sudden modification of the photon dispersion relation in the waveguide - a quantum analogue of the emission of cosmological radiation in a curved space for spontaneous particle pair creations \cite{lang2018,Tian_2017} also refereed as dynamical Casimir effect \cite{wilson2011observation}. The symmetry ensures that collective excitations of the array decay into the entangled microwave beams propagating along the waveguide in the opposite directions. 

Compared to the prior art, the proposed mechanism does not involve the use of an external magnetic field \cite{Laetheenmaeki_2013} or a pump field within a waveguide \cite{grimsmo2017squeezing}. It predicts quantum correlations at distance and therefore differs  from other studies like the two photons correlations analyzed in \cite{FANG201692,PhysRevA.96.013842,PhysRevA.91.053845}, 
or sub- and superradiance in \cite{PhysRevA.88.043806} or even phase transition \cite{zhang2014quantum}. 

\begin{figure}
 \centering
 \includegraphics[width=8.5cm]{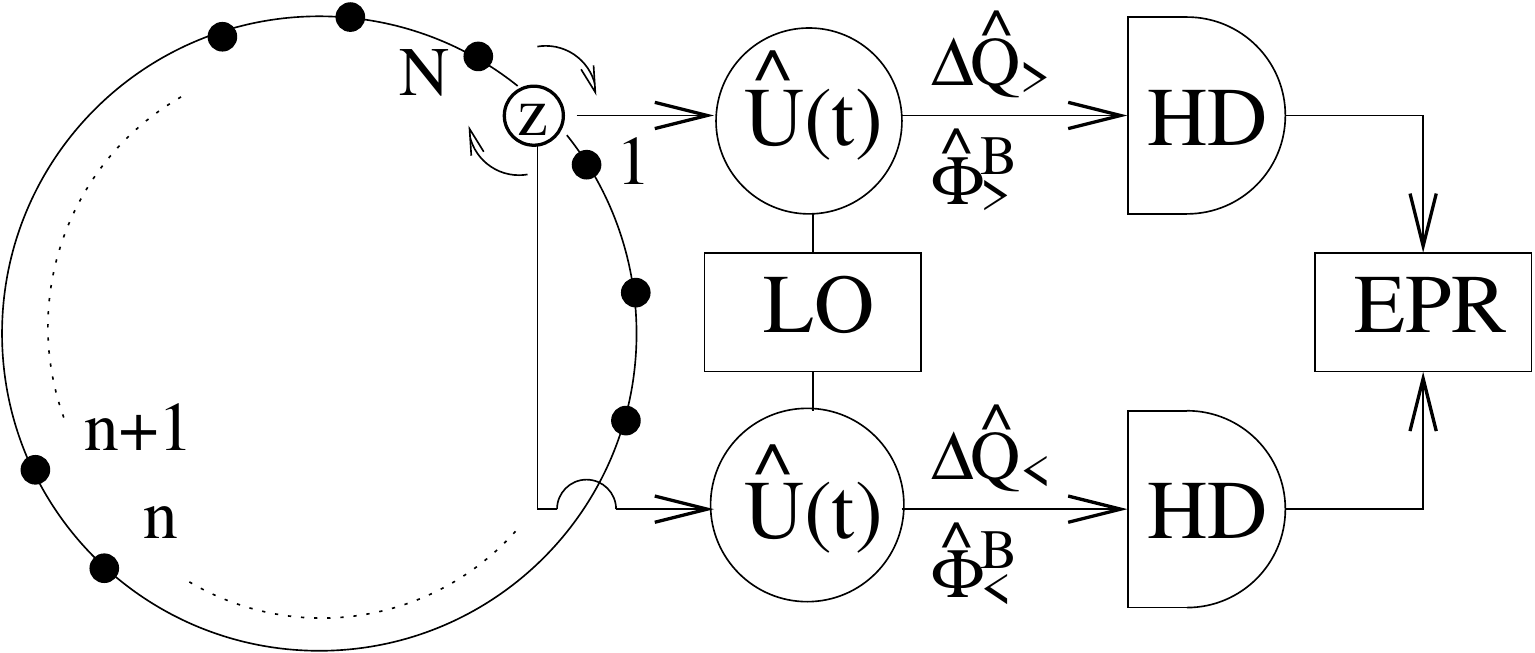}
 \caption{$N$ qu-bits are imbedded in a ring waveguide. The emitted signal is collected in a circulator $z$. One of its quadratures (charge  $\Delta \hat Q_\gtrless $ or flux $\hat \Phi^B_\gtrless$) or their combination  is subject to a homodyne detection (HD) after a mixing with a local oscillator (LO) described effectively by  a unitary operator $\hat U(t)$. The resultsing EPR correlation between the output signals is detected. }
 \label{fig:1}
\end{figure}

{\it Setup:}
The proposed scheme is presented in Fig.1 and Fig.2. An array of $N$ transmon qubits is embedded in a ring waveguide at zero temperature. Starting at equilibrium, we adiabatically increase the potential of the qubit island to $V$, and then suddenly drop it to zero. The initial charge on the island is $q_n =C_0 V$, where $C_0$ is the effective  capacitance. Classically, the island charge oscillates at a frequency 
$\epsilon_0 =\sqrt{8E_J e^2 /C_0}/\hbar$, where Josephson energy $E_J \gg e^2/C_0$ for a transmon. In the quantum case these oscillations will decay into the electromagnetic vacuum modes by producing two counterpropagating entangled beams. 
The subsequent action of a circulator passes  these 
outputs on for a homodyne detection \cite{HD}, that includes local oscillator mixing and frequency filtering,  
in order to determine their mutual EPR-type quantum correlations \cite{PhysRevA.65.013813,PhysRev.47.777}.  
The integral of the voltage pulse (flux \cite{zagoskin2011quantum}) and the total induced charge correspond to correlated (resp. anticorrelated) continuous variable quadratures.


\begin{figure}
\begin{center}
 \includegraphics[width=9cm]{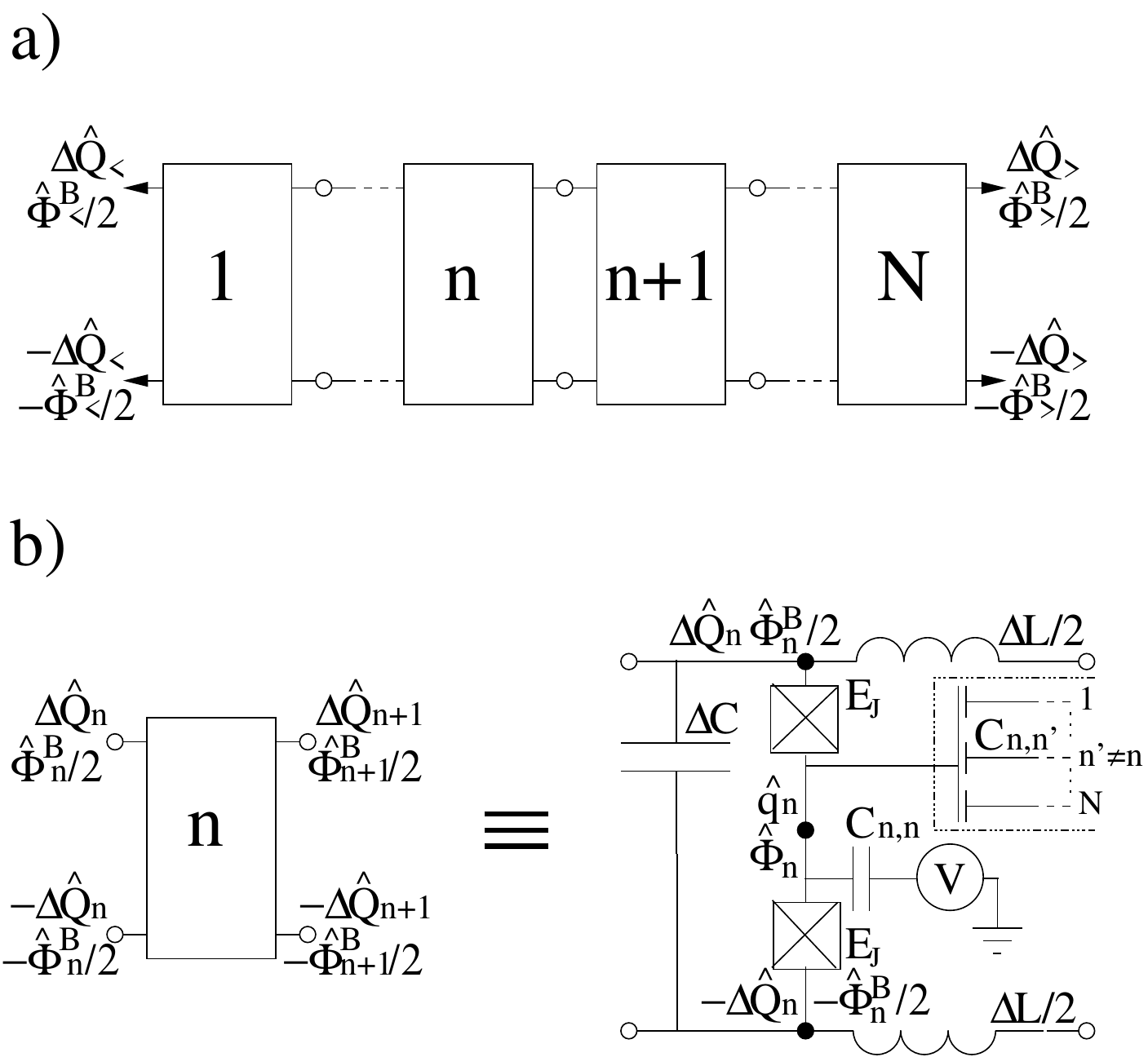}
\caption{Quantum circuit corresponding to the Hamiltonian (\ref{chargephase}). (a) The ring waveguide cut at the circulator is represented by an array of transmons ($n = 1,2.. N$) coupled to the microwave radiation mode. The charges $\pm \Delta \hat Q_{\gtrless}$ and  fluxes  $\pm \hat \Phi_{\gtrless}^{B}/2$ are the output components for a homodyne detection. (b) Transmon qubit embedded in a waveguide. Each node (fat black dot) is associated with 
the excess charge $\hat q_n$ and  flux 
$\hat \Phi_{n}$ operators describing the transmon, and the incremental charge $\pm \Delta \hat Q_n$ and  flux 
$\pm \hat \Phi_{n}^{B}/2$ operators describing the EM field within the waveguide.  
}
 \end{center}
\end{figure}

{\it Theoretical model:} The lumped-elements scheme of the device is shown in Fig.2. The $n$th transmon's quantum operators are its excess charge $\hat q_n =(2e)(\hat n_n - N_s)$ measured from the equilibrium value $N_s$ and  flux $\hat \Phi_{n}$. Its mutual capacitance with the neutral gate (with gate voltage $V$) is $C_{n,n}$, and $C_{n,n'}$ is the mutual capacitance between the $n$th and $n'$th transmons. We omit the capacitive couplings to the waveguide as they cancel out for the decay process. 
The waveguide is described by 
conjugated operators charges $\Delta \hat Q_n$ and fluxes $\hat \Phi_{n}^{B}$, and is characterized by the mutual capacitance $\Delta C$ and inductance 
$\Delta L/2$ between two adjacent transmons. These operators obey 
the canonical 
commutation relations $[\hat \Phi_n,\hat q_{n'}]=i\hbar\delta_{n,n'}$ and 
$[ \hat \Phi_{n}^{B}, \Delta \hat Q_{n'}]=i\hbar\delta_{n,n'}$.

The Hamiltonian of the system is obtained by quantizing the different components' contributions to the total energy  \cite{vool2017introduction}, yielding:
\begin{widetext}
\begin{eqnarray}\label{chargephase}
\! \! \! \hat H
= \! \sum_{n=1}^N 
V \hat q_n +\!
\sum_{n'=1}^N \frac{C^{-1}_{n,n'}}{2}
 \hat q_{n'} \hat q_n
-E_J^b\! \left[
\cos \!\left[\frac{2e}{\hbar}\left(\hat \Phi_n -\frac{\hat \Phi_{n}^{B}}{2}\right)\right] +
\cos \! \left[\frac{2e}{\hbar}\left(\hat \Phi_n +\frac{\hat \Phi_{n}^{B}}{2}\right)\right]\right] \! 
+\frac{\Delta \hat Q_n^2 }{2\Delta C} + 
\frac{(\hat \Phi_{n+1}^{B}- \hat \Phi_{n}^{B})^2}{2\Delta L}
\end{eqnarray}
\end{widetext}
where $E_J^b$ is the bare Josephson energy of a junction (see Fig.2). The effective 
renormalized Josephson energy $E_J= E_{J}^b\langle 0 |\cos\left(2e\hat \Phi_n/\hbar \right)
\cos\left(e\hat \Phi^B_n/\hbar \right)|0\rangle$  is defined with respect  to the vacuum energy state $|0\rangle$.
The essentially nonlinear cosine interaction terms between the transmons and the electromagnetic modes result from the Josephson junctions and have been configured so as to be an even function of each field amplitude. 

The transmons can be close enough to each other to be  
coupled through the mutual capacitances. Assuming 
the translational invariance of the ring, $C_{n,n'}$ depends only on the distance $n-n'$ modulo $N$ and it is convenient to define their Fourier components: 
$C_k=\sum_{n=1}^N e^{-i2\pi k (n-n')/N} C_{n,n'}/N$. Here the integer $k$ is defined modulo $N$, and the capacitance energy can be written as
$E_{C,k}=(2e)^2/2C_k$. Under these conditions, 
we can rewrite the transmon operators in their ''wavevector'' components as:
\begin{eqnarray}
\hat \Phi_n=\frac{\hbar}{2e} \sum_{k=1}^N \left(\frac{E_{C,k}}{ E_J}\right)^{1/4}
 \frac{e^{i \frac{2\pi k n}{N}} (\hat b_k+\hat b^\dagger_{-k})}{\sqrt{2N}}
\\
\hat q_{n}= (2e)
\sum_{k=1}^N \left(\frac{E_{C,k}}{E_J}\right)^{-1/4}
\frac{e^{i \frac{2\pi k n}{N}} (\hat b_k-\hat b^\dagger_{-k})}{i\sqrt{2N}}\, .
\end{eqnarray}
The creation-annihilation operators $\hat b^\dagger_k$ and $\hat b_{k}$ describe plasmon-like collective excitations of 
charge motion with a wavenumber given by $K=2\pi k/N$.

The charge and flux operators can be similarly defined through the electromagnetic field component as 
$\hat \Phi_{n}^{B}=\hat{\alpha}_n/\sqrt{\Delta C}$ and  
$\Delta \hat Q_n=\sqrt{\Delta C}
\dot {\hat{\alpha}}_n$. Here $\hat{\alpha}_n$ is the vector potential for an ideal waveguide consisting of 
two parallel infinite planes \cite{rakhmanov2008quantum}. 
It can be expressed through the wavevector components: 
\begin{eqnarray}
\hat \alpha_n  = \sum_{k=1}^N e^{i 2\pi k n/N}\sqrt{\frac{\hbar}{2 \omega_k N}}( \hat a_k +  \hat a_{-k}^\dagger )
\\
{\dot { \hat\alpha}}_n= \sum_{k=1}^N e^{2\pi i k n/N} \sqrt{\frac{\hbar\omega_k}{2N}}( \hat a_k - \hat a_{-k}^\dagger)/i
\end{eqnarray}
where $\hat a^\dagger_k$ and $\hat a_{k}$ are the photon creation-annihilation operators.
The vacuum state  is, as usual, defined from  $\hat a_k|0 \rangle=0$ and $\hat b_k|0 \rangle=0$. 
With respect to this vacuum definition and in the weak coupling approximation, the Hamiltonian (\ref{chargephase}) is rewritten for 
$V=0$ in terms of the circuit component characteristics only, up to fourth order in the field and in a normal ordered form as:
\begin{widetext}
\begin{eqnarray}\label{bfield}
\!\!\! :\hat H:
&=& \sum_{k=1}^N \hbar \epsilon_k \hat b^\dagger_{k}\hat b_k+\hbar \omega_k \hat a_k^\dagger  \hat a_k 
+ 
\frac{E_{\Delta C}}{32 N} \sum_{k',l=1}^N
\sqrt{\frac{\epsilon_{k+l} \epsilon_{k'-l}}{\omega_k \omega_{k'}}}
:(\hat b^\dagger_{k+l} + \hat b_{-k-l})(\hat b^\dagger_{k'-l} + \hat b_{-k'+l})
(\hat a_k +  \hat a_{-k}^\dagger)(\hat a_{k'} +  \hat a_{-k'}^\dagger ):
\end{eqnarray}
\end{widetext}
The first and second terms correspond respectively  to the plasmon mode with 
energy spectrum $\hbar \epsilon_k= \sqrt{4E_JE_{C,k}}$ and 
the photon mode with $\omega_{k}=\sqrt{2(1-\cos(K))/(\Delta C \Delta L) +\omega_0^2}$, where 
$\hbar \omega_0=\sqrt{E_JE_{\Delta C}}$ and $E_{\Delta C}=2e^2 /(\hbar^2\Delta C)$. 
The plasmon spectrum is almost flat. The photon spectrum has a gap resulting from the Josephson energy contribution. 
Without it, the spectrum would be linear with a light speed 
$c =D/\sqrt{\Delta C \Delta L}$ where $D$ is the transmon interdistance. The third term is the quartic interaction 
responsible of for the coupling   $k + k' \leftrightarrow (k+l) + (k'-l)$ between the  radiation and the transmon qubits and is negligible  only if $E_{\Delta C} \ll \epsilon_k, \omega_k $.
Note that we neglect the quartic self-modulation  terms 
responsible for anharmonicity \cite{PhysRevA.76.042319} for both fields since these are even weaker than the coupling.

{\it The proposed experiment:}
We start by adiabatically applying to each qubit the potential $V$, producing the initial charge with the non zero expectation $\langle \hat q_n \rangle=C_0V$ interpreted as displacement of the vacuum state. 
Then the potential is suddenly dropped to zero. The qubit islands begin discharging, emitting in the process entangled pairs of photons. 
The corresponding transmon state is a coherent state with $k=0$. Its amplitude at $t=0$ is                                                          
\begin{eqnarray}
\varphi_0 \equiv \sqrt{N}\langle \hat b_0 \rangle = i \sqrt{E_J/\epsilon_0}(eV/E_{C,0}).
\end{eqnarray}
The qu-bit regime is recovered in the case of fainted coherent state. 
If parametrize the squeezed radiation mode with  
the squeezing amplitude $r_k(t)$ and the phase $\theta_k(t)$, the full Ansatz for the quantum state of radiation in the waveguide is 
\footnote{See the Supplemental Materials for the mathematical details}
\begin{eqnarray}
|\Psi(t)\rangle= \prod_{k=1}^{N/2}
e^{ r_k (e^{-2i\theta_k}\hat a_{k}^\dagger \hat a^\dagger_{-k}
-e^{2i\theta_k} \hat a_{k}  \hat a_{-k})}  
\hat D(t)|0\rangle,
\end{eqnarray}
with the displacement unitary 
transformation $
\hat D(t) =\exp(\sqrt{N}\varphi\hat b_0^\dagger -\sqrt{N}\varphi^* \hat b_0)$.

From the Lagrangian $\Re[\langle \Psi(t)|(i\hbar)
\partial_t -:\hat H:|\Psi(t)\rangle] $ we
obtain the dynamical equations:
\begin{eqnarray} \label{phi}
i\dot{ \varphi}&=&  \epsilon_0  \varphi
+ 
\frac{E_{\Delta C}\epsilon_0}{8N}(\varphi + \varphi^*)
\nonumber \\
&\times &
\sum_{k=1}^{N/2} 
\frac{\cosh(2r_k)-1+\cos(2\theta_k)\sinh(2r_k) }{\hbar\omega_k}
\\ \label{theta} 
\dot {\theta}_k
\! &=&\!
 \omega_k \!
+ \! \frac{E_{\Delta C}\epsilon_0}{16}
(\varphi + \varphi^*)^2 
\frac{1+\cos(2\theta_k)\!\coth(2r_k) }{\hbar\omega_k}
\\ \label{r} 
\dot {r}_k
&=& \frac{E_{\Delta C}\epsilon_0}{16}(\varphi +  \varphi^*)^2
\frac{\sin(2\theta_k)}{\hbar \omega_k}.
\end{eqnarray}

In the short time limit, assuming that all the capacitances are of the same order of magnitude and taking realistic values for the system parameters ($E_{C,0} \sim E_{\Delta C} \sim 10 {\rm GHz}$,  $E_J \sim 1{\rm THz}$), we can make the following direct estimates.  The rate of squeezing is  
${\dot r}_k(0) \sim  E_J (eV/ E_{\Delta C})^2/\hbar 
\sim 1 {\rm GHz}$  for the initial voltage $V= 10 \mu$V.  
The relative charge leakage, $\langle \Delta q_n \rangle/
\langle q_n \rangle \sim \sqrt{E_J/E_{C,0}} (eVt/\hbar)^2 \sim t^2({\rm ns}^2)$, is quadratic in time.

In the long time limit, the equations are solved in the 
rotating wave approximation. Then the decaying of two transmon excitations into 
two photons satisfies the number and energy conservation $2\epsilon_0=2\omega_k$. 
We consider two distinct cases of a transmon 
excitation decaying into either a continuum of photon modes or into 
a single mode.

{\it Decay into a continuum :}
In the large-$N$ limit and for weak squeezing $r_k(t) \leq 1$, we can approximate the phase by 
$\theta_k(t)=\hbar\omega_kt-\pi/4$. The solution for the transmon field in the continuum limit is then
$\varphi (t) = ie^{-i\epsilon_0 t}
\varphi_0/\sqrt{1+ \Gamma t}$
with a inverse power decay rate:
\begin{eqnarray}
\!\!\Gamma = \left(\frac{E_{\Delta C} eV }{16 E_{C,0}}\right)^2 \frac{E_J/\hbar^3 }{
\sqrt{(\omega^2_{\frac{N}{2}}-\epsilon_0^2)(\epsilon_0^2-\omega^2_{0})}}\sim
\frac{C_0V^2}{ \hbar}
\end{eqnarray}
This rate corresponds to the capacitance energy perturbation introduced initially and has to be much less than the plasmon frequency
$\epsilon_0$ (typically in the GHz-THz range) but 
much larger than any decoherence rate (in the MHz range)\cite{PhysRevLett.107.240501}. 
For the squeezing parameter, we obtain:
\begin{eqnarray}
r_k(t)
&=&\frac{E_{\Delta C}}{16} \int_0^t dt' 
\frac{|\varphi (t')|^2 \epsilon_0\cos[2\delta_k t']}
{\hbar^2\omega_k} 
\end{eqnarray}
where $\delta_k=\omega_k-
\epsilon_0$ is the detuning frequency. 
Fig.3 represents its growing with time and
concentrating at zero detuning.
\begin{figure}
 \begin{center}
 \includegraphics[width=8.5cm]{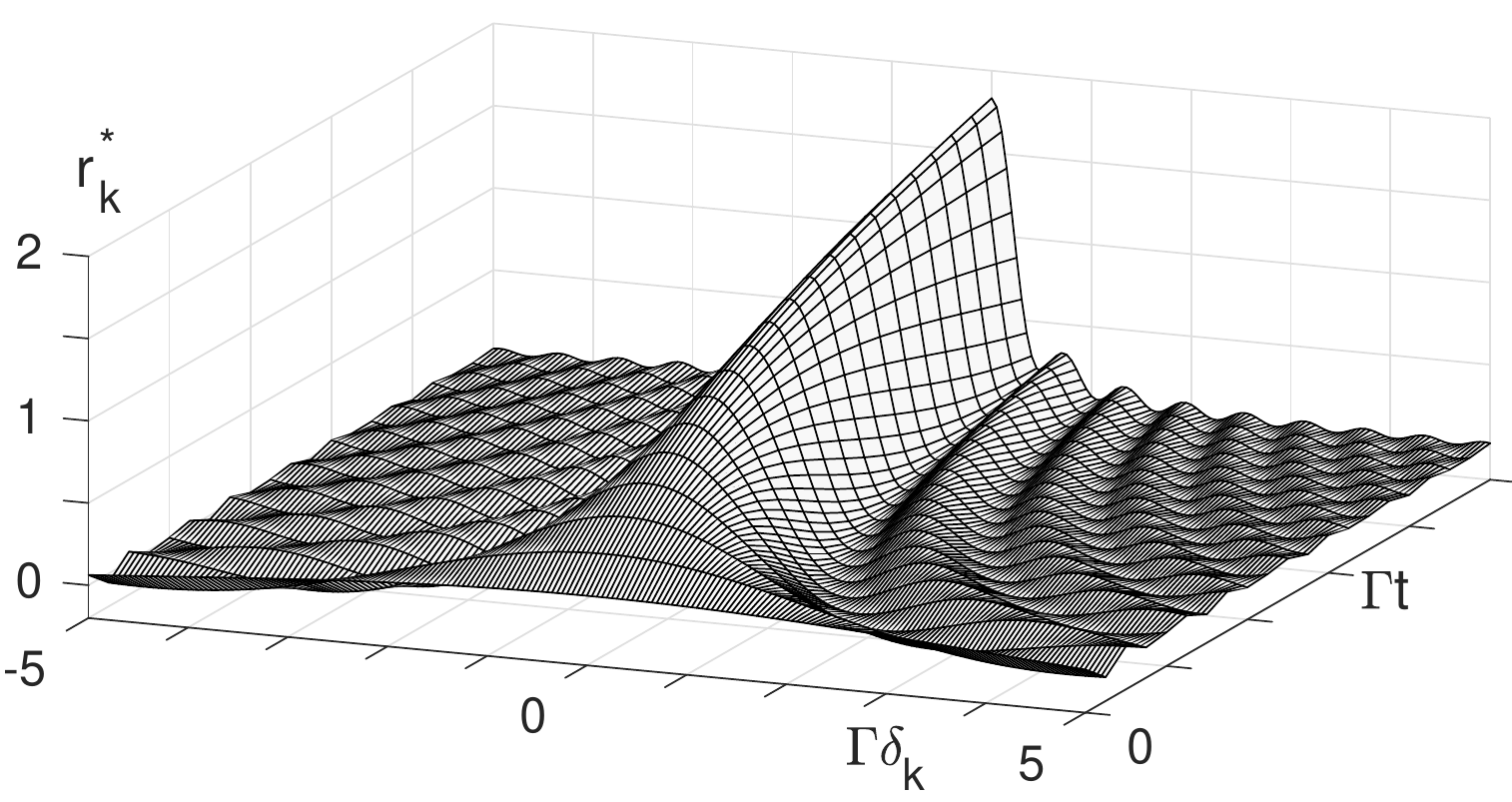}
 \end{center}
 \caption{Dimensionless squeezing parameter $r^*_k(t)=(16\hbar\Gamma/E_{\Delta C})(r_k(t)/|\phi_0|^2)$ as a function of time and detuning frequency for $\epsilon_0\gg \Gamma$.}
 \label{fig:3}
\end{figure}

The total number of photons in one direction $k\gtrless 0$ can then be estimated as
\begin{eqnarray}
\langle \hat N_{ph}^\gtrless \rangle= 
\sum_{k=1}^{N/2} \sinh^2(r_k(t))=N\frac{|\varphi_0|^2-|\varphi (t)|^2}{2}
\end{eqnarray}
The photon waves along the two opposite direction have the perfect entanglement correlation 
$\langle (\delta \hat N_{ph}^> -\delta \hat N_{ph}^< )^2\rangle =0$.

Their quadratures are also correlated and are measured through a homodyne detection after mixing them with a local oscillator of frequency $\omega$
that is simply described via the unitary operator
$\hat U(t)=e^{i(\omega t-\pi/4)\sum_{k=1}^{N} \hat a^\dagger_k\hat a_k}$. The effective output signals are
\begin{eqnarray}
\!\! \hat \alpha_n^{out} =\hat U^\dagger(t)\hat \alpha_n \hat U(t) 
\quad \quad 
{\dot { \hat\alpha}}^{out}_n=
\hat U^\dagger (t){\dot {\hat \alpha}}_n \hat U(t)
\end{eqnarray}
Their averages are zero. However, 
we determine EPR correlations for the Fourier components
$\hat \alpha_k^{out}$,  $\dot{ \hat \alpha}_k^{out}$ of these 
continuous variables relatively to the shot noise level 
\cite{PhysRevA.65.013813}:
\begin{eqnarray}
\frac{\langle ( \hat \alpha_k^{out} - \hat \alpha_{-k}^{out})^2 \rangle}{2\langle {\hat \alpha}_k^{out\, 2}\rangle|_{r_k=0}}=
\frac{\langle ( \dot{ \hat \alpha}_k^{out} + \dot{\hat \alpha}_{-k}^{out})^2\rangle }{2\langle \dot{\hat \alpha}_k^{out\, 2}\rangle |_{r_k=0}}
\stackrel{\omega \rightarrow \omega_k}{=}e^{-2 r_k(t)}
\nonumber  \\
\end{eqnarray}
These correlations become important for large squeezing.  
The corresponding physical quadratures are the  charge 
$\Delta \hat Q_\gtrless = 
\sqrt{\Delta C}
{\dot {\hat \alpha}}^{out}_{\pm k}$ and the flux  
$\hat \Phi^B_\gtrless= \hat \alpha^{out}_{\pm k}/ \sqrt{\Delta C}$ in the waveguide, which are, respectively, anticorrelated and correlated. 

{\it Oscillation with two modes:}
In case of long wavelengths (GHz), the frequency separation between modes in the ring becomes large. We can then select only two entangled modes $\pm k$ only in the Eqs.(\ref{phi},\ref{theta},\ref{r}), which interact with a resonant transmon mode. Other photon modes are not perturbed.

Besides the interaction terms for the transition,
an additional modulation phase term 
affects the transition frequency \cite{grimsmo2017squeezing}. 
The maximum squeeezing that can be reached is  
$r_{m}\stackrel{{ N}\rightarrow \infty}{=}\ln(2 N |\varphi_0|^2)/2$
corresponding to  the total depletion $|\varphi(t)|^2=0$. 
For simplicity, we shall assume the phase modulation term is constant which implies the restriction to values 
$r_k(t) \leq r_m- \sqrt{N/2} e^{-r_m}$.
We define the dimensionless parameters:
$\tilde t= t E_{\Delta C}\epsilon_0 /(32\hbar\omega_k)$ and 
$\tilde \delta = \hbar \delta_k 32 \omega_k /E_{\Delta C}\epsilon_0$. 
Two cases are considered:

{\it 1) No phase modulation:}
For short times, we note that the fastest squeezing rate is achieved if 
the phase matching condition ${\tilde \delta}=4|\varphi_0|^2$ is satisfied. 
Using this condition, the phase modulation can be neglected and the squeezing parameter evolves towards $r_m$.
The explicit expression is
\begin{eqnarray}
r_k(t) =\frac{1}{2}\ln\left(\frac{e^{2r_{m}}+e^{-4\sinh(2r_{m}){\tilde t}/N}}
{1+e^{2r_{m}}e^{-4\sinh(2r_{m}){\tilde t}/N}}\right)
\end{eqnarray}

{\it 2) Weak depletion:}
When the the detuning is not phase matched, 
the charge leakage from the island can be neglected, i.e $|\phi(t)| \simeq |\phi_0|$. 
For a small value of detuning within the interval $ -2|\varphi_0|^2 \leq {\tilde \delta}  \leq 6|\varphi_0|^2$, 
the photon number grows exponentially:
$N_{ph}(t)=\sinh^2(r_k(t))=8|\varphi_0|^2\sinh^2(\Omega {\tilde t})/\Omega^2$ with the characteristic angular frequency 
$
\Omega=\sqrt{|({\tilde \delta} -4|\varphi_0|^2)^2 -4|\varphi_0|^4|}$.
Outside this interval, 
the solution becomes $N_{ph}(t)=8|\varphi_0|^2\sin^2(\Omega {\tilde t})/\Omega^2$, which corresponds to a  Rabi-like oscillation between the plasmon mode and the photon modes. This  Rabi-like superposition of a plasmon state and an EPR photon state illustrates  
the  rich possibilities offered by this qubit line device 
for quantum design. 

{\it Conclusions:}
We propose the superconducting transmon line embedded 
in a ring waveguide as a generator of entangled beams  of microwave radiation. Using the fully quantum description, we 
can describe the scattering  process between  photons and the collective transmon excitation. 
We also show that high squeezing may be obtained in the long wavelength regime, allowing for a genuine EPR-like experiment in 
a microchip device. An interesting extension of this design would be a parametric optical amplifier with the proposed setup for quantum imaging \cite{PhysRevA.65.013813,grimsmo2017squeezing}.

{\bf Acknowledgements:}
PN thanks G. Tsironis, Z. Ivic and J. Brehm for helpful discussions and the Dept.
of Mathematics and Statistics, University of Saskatchewan, for
hospitality. AZ was partially supported by the NDIAS Residential Fellowship (University of Notre Dame).

\bibliographystyle{apsrev4-1}
\bibliography{paperrefs}

\begin{thebibliography}{32}%
\makeatletter
\providecommand \@ifxundefined [1]{%
 \@ifx{#1\undefined}
}%
\providecommand \@ifnum [1]{%
 \ifnum #1\expandafter \@firstoftwo
 \else \expandafter \@secondoftwo
 \fi
}%
\providecommand \@ifx [1]{%
 \ifx #1\expandafter \@firstoftwo
 \else \expandafter \@secondoftwo
 \fi
}%
\providecommand \natexlab [1]{#1}%
\providecommand \enquote  [1]{``#1''}%
\providecommand \bibnamefont  [1]{#1}%
\providecommand \bibfnamefont [1]{#1}%
\providecommand \citenamefont [1]{#1}%
\providecommand \href@noop [0]{\@secondoftwo}%
\providecommand \href [0]{\begingroup \@sanitize@url \@href}%
\providecommand \@href[1]{\@@startlink{#1}\@@href}%
\providecommand \@@href[1]{\endgroup#1\@@endlink}%
\providecommand \@sanitize@url [0]{\catcode `\\12\catcode `\$12\catcode
  `\&12\catcode `\#12\catcode `\^12\catcode `\_12\catcode `\%12\relax}%
\providecommand \@@startlink[1]{}%
\providecommand \@@endlink[0]{}%
\providecommand \url  [0]{\begingroup\@sanitize@url \@url }%
\providecommand \@url [1]{\endgroup\@href {#1}{\urlprefix }}%
\providecommand \urlprefix  [0]{URL }%
\providecommand \Eprint [0]{\href }%
\providecommand \doibase [0]{http://dx.doi.org/}%
\providecommand \selectlanguage [0]{\@gobble}%
\providecommand \bibinfo  [0]{\@secondoftwo}%
\providecommand \bibfield  [0]{\@secondoftwo}%
\providecommand \translation [1]{[#1]}%
\providecommand \BibitemOpen [0]{}%
\providecommand \bibitemStop [0]{}%
\providecommand \bibitemNoStop [0]{.\EOS\space}%
\providecommand \EOS [0]{\spacefactor3000\relax}%
\providecommand \BibitemShut  [1]{\csname bibitem#1\endcsname}%
\let\auto@bib@innerbib\@empty
\bibitem [{\citenamefont {Paik}\ \emph {et~al.}(2011)\citenamefont {Paik},
  \citenamefont {Schuster}, \citenamefont {Bishop}, \citenamefont {Kirchmair},
  \citenamefont {Catelani}, \citenamefont {Sears}, \citenamefont {Johnson},
  \citenamefont {Reagor}, \citenamefont {Frunzio}, \citenamefont {Glazman},
  \citenamefont {Girvin}, \citenamefont {Devoret},\ and\ \citenamefont
  {Schoelkopf}}]{PhysRevLett.107.240501}%
  \BibitemOpen
  \bibfield  {author} {\bibinfo {author} {\bibfnamefont {H.}~\bibnamefont
  {Paik}}, \bibinfo {author} {\bibfnamefont {D.~I.}\ \bibnamefont {Schuster}},
  \bibinfo {author} {\bibfnamefont {L.~S.}\ \bibnamefont {Bishop}}, \bibinfo
  {author} {\bibfnamefont {G.}~\bibnamefont {Kirchmair}}, \bibinfo {author}
  {\bibfnamefont {G.}~\bibnamefont {Catelani}}, \bibinfo {author}
  {\bibfnamefont {A.~P.}\ \bibnamefont {Sears}}, \bibinfo {author}
  {\bibfnamefont {B.~R.}\ \bibnamefont {Johnson}}, \bibinfo {author}
  {\bibfnamefont {M.~J.}\ \bibnamefont {Reagor}}, \bibinfo {author}
  {\bibfnamefont {L.}~\bibnamefont {Frunzio}}, \bibinfo {author} {\bibfnamefont
  {L.~I.}\ \bibnamefont {Glazman}}, \bibinfo {author} {\bibfnamefont {S.~M.}\
  \bibnamefont {Girvin}}, \bibinfo {author} {\bibfnamefont {M.~H.}\
  \bibnamefont {Devoret}}, \ and\ \bibinfo {author} {\bibfnamefont {R.~J.}\
  \bibnamefont {Schoelkopf}},\ }\href {\doibase 10.1103/PhysRevLett.107.240501}
  {\bibfield  {journal} {\bibinfo  {journal} {Phys. Rev. Lett.}\ }\textbf
  {\bibinfo {volume} {107}},\ \bibinfo {pages} {240501} (\bibinfo {year}
  {2011})}\BibitemShut {NoStop}%
\bibitem [{\citenamefont {Koch}\ \emph {et~al.}(2007)\citenamefont {Koch},
  \citenamefont {Yu}, \citenamefont {Gambetta}, \citenamefont {Houck},
  \citenamefont {Schuster}, \citenamefont {Majer}, \citenamefont {Blais},
  \citenamefont {Devoret}, \citenamefont {Girvin},\ and\ \citenamefont
  {Schoelkopf}}]{PhysRevA.76.042319}%
  \BibitemOpen
  \bibfield  {author} {\bibinfo {author} {\bibfnamefont {J.}~\bibnamefont
  {Koch}}, \bibinfo {author} {\bibfnamefont {T.~M.}\ \bibnamefont {Yu}},
  \bibinfo {author} {\bibfnamefont {J.}~\bibnamefont {Gambetta}}, \bibinfo
  {author} {\bibfnamefont {A.~A.}\ \bibnamefont {Houck}}, \bibinfo {author}
  {\bibfnamefont {D.~I.}\ \bibnamefont {Schuster}}, \bibinfo {author}
  {\bibfnamefont {J.}~\bibnamefont {Majer}}, \bibinfo {author} {\bibfnamefont
  {A.}~\bibnamefont {Blais}}, \bibinfo {author} {\bibfnamefont {M.~H.}\
  \bibnamefont {Devoret}}, \bibinfo {author} {\bibfnamefont {S.~M.}\
  \bibnamefont {Girvin}}, \ and\ \bibinfo {author} {\bibfnamefont {R.~J.}\
  \bibnamefont {Schoelkopf}},\ }\href {\doibase 10.1103/PhysRevA.76.042319}
  {\bibfield  {journal} {\bibinfo  {journal} {Phys. Rev. A}\ }\textbf {\bibinfo
  {volume} {76}},\ \bibinfo {pages} {042319} (\bibinfo {year}
  {2007})}\BibitemShut {NoStop}%
\bibitem [{\citenamefont {Wendin}(2017)}]{wendin2017quantum}%
  \BibitemOpen
  \bibfield  {author} {\bibinfo {author} {\bibfnamefont {G.}~\bibnamefont
  {Wendin}},\ }\href@noop {} {\bibfield  {journal} {\bibinfo  {journal}
  {Reports on Progress in Physics}\ }\textbf {\bibinfo {volume} {80}},\
  \bibinfo {pages} {106001} (\bibinfo {year} {2017})}\BibitemShut {NoStop}%
\bibitem [{\citenamefont {Makhlin}\ \emph {et~al.}(2001)\citenamefont
  {Makhlin}, \citenamefont {Sch\"on},\ and\ \citenamefont
  {Shnirman}}]{RevModPhys.73.357}%
  \BibitemOpen
  \bibfield  {author} {\bibinfo {author} {\bibfnamefont {Y.}~\bibnamefont
  {Makhlin}}, \bibinfo {author} {\bibfnamefont {G.}~\bibnamefont {Sch\"on}}, \
  and\ \bibinfo {author} {\bibfnamefont {A.}~\bibnamefont {Shnirman}},\ }\href
  {\doibase 10.1103/RevModPhys.73.357} {\bibfield  {journal} {\bibinfo
  {journal} {Rev. Mod. Phys.}\ }\textbf {\bibinfo {volume} {73}},\ \bibinfo
  {pages} {357} (\bibinfo {year} {2001})}\BibitemShut {NoStop}%
\bibitem [{\citenamefont {Zagoskin}\ \emph {et~al.}(2014)\citenamefont
  {Zagoskin}, \citenamefont {Il'ichev}, \citenamefont {Grajcar}, \citenamefont
  {Betouras},\ and\ \citenamefont {Nori}}]{zagoskin2014test}%
  \BibitemOpen
  \bibfield  {author} {\bibinfo {author} {\bibfnamefont {A.~M.}\ \bibnamefont
  {Zagoskin}}, \bibinfo {author} {\bibfnamefont {E.}~\bibnamefont {Il'ichev}},
  \bibinfo {author} {\bibfnamefont {M.}~\bibnamefont {Grajcar}}, \bibinfo
  {author} {\bibfnamefont {J.~J.}\ \bibnamefont {Betouras}}, \ and\ \bibinfo
  {author} {\bibfnamefont {F.}~\bibnamefont {Nori}},\ }\href@noop {} {\bibfield
   {journal} {\bibinfo  {journal} {Frontiers in Physics}\ }\textbf {\bibinfo
  {volume} {2}},\ \bibinfo {pages} {33} (\bibinfo {year} {2014})}\BibitemShut
  {NoStop}%
\bibitem [{\citenamefont {Navez}\ \emph {et~al.}(2017)\citenamefont {Navez},
  \citenamefont {Tsironis},\ and\ \citenamefont
  {Zagoskin}}]{PhysRevB.95.064304}%
  \BibitemOpen
  \bibfield  {author} {\bibinfo {author} {\bibfnamefont {P.}~\bibnamefont
  {Navez}}, \bibinfo {author} {\bibfnamefont {G.~P.}\ \bibnamefont {Tsironis}},
  \ and\ \bibinfo {author} {\bibfnamefont {A.~M.}\ \bibnamefont {Zagoskin}},\
  }\href {\doibase 10.1103/PhysRevB.95.064304} {\bibfield  {journal} {\bibinfo
  {journal} {Phys. Rev. B}\ }\textbf {\bibinfo {volume} {95}},\ \bibinfo
  {pages} {064304} (\bibinfo {year} {2017})}\BibitemShut {NoStop}%
\bibitem [{\citenamefont {Wallraff}\ \emph {et~al.}(2004)\citenamefont
  {Wallraff}, \citenamefont {Schuster}, \citenamefont {Blais}, \citenamefont
  {Frunzio}, \citenamefont {Huang}, \citenamefont {Majer}, \citenamefont
  {Kumar}, \citenamefont {Girvin},\ and\ \citenamefont
  {Schoelkopf}}]{wallraff2004strong}%
  \BibitemOpen
  \bibfield  {author} {\bibinfo {author} {\bibfnamefont {A.}~\bibnamefont
  {Wallraff}}, \bibinfo {author} {\bibfnamefont {D.~I.}\ \bibnamefont
  {Schuster}}, \bibinfo {author} {\bibfnamefont {A.}~\bibnamefont {Blais}},
  \bibinfo {author} {\bibfnamefont {L.}~\bibnamefont {Frunzio}}, \bibinfo
  {author} {\bibfnamefont {R.-S.}\ \bibnamefont {Huang}}, \bibinfo {author}
  {\bibfnamefont {J.}~\bibnamefont {Majer}}, \bibinfo {author} {\bibfnamefont
  {S.}~\bibnamefont {Kumar}}, \bibinfo {author} {\bibfnamefont {S.~M.}\
  \bibnamefont {Girvin}}, \ and\ \bibinfo {author} {\bibfnamefont {R.~J.}\
  \bibnamefont {Schoelkopf}},\ }\href@noop {} {\bibfield  {journal} {\bibinfo
  {journal} {Nature}\ }\textbf {\bibinfo {volume} {431}},\ \bibinfo {pages}
  {162} (\bibinfo {year} {2004})}\BibitemShut {NoStop}%
\bibitem [{\citenamefont {Il'ichev}\ \emph {et~al.}(2003)\citenamefont
  {Il'ichev}, \citenamefont {Oukhanski}, \citenamefont {Izmalkov},
  \citenamefont {Wagner}, \citenamefont {Grajcar}, \citenamefont {Meyer},
  \citenamefont {Smirnov}, \citenamefont {Maassen van~den Brink}, \citenamefont
  {Amin},\ and\ \citenamefont {Zagoskin}}]{PhysRevLett.91.097906}%
  \BibitemOpen
  \bibfield  {author} {\bibinfo {author} {\bibfnamefont {E.}~\bibnamefont
  {Il'ichev}}, \bibinfo {author} {\bibfnamefont {N.}~\bibnamefont {Oukhanski}},
  \bibinfo {author} {\bibfnamefont {A.}~\bibnamefont {Izmalkov}}, \bibinfo
  {author} {\bibfnamefont {T.}~\bibnamefont {Wagner}}, \bibinfo {author}
  {\bibfnamefont {M.}~\bibnamefont {Grajcar}}, \bibinfo {author} {\bibfnamefont
  {H.-G.}\ \bibnamefont {Meyer}}, \bibinfo {author} {\bibfnamefont {A.~Y.}\
  \bibnamefont {Smirnov}}, \bibinfo {author} {\bibfnamefont {A.}~\bibnamefont
  {Maassen van~den Brink}}, \bibinfo {author} {\bibfnamefont {M.~H.~S.}\
  \bibnamefont {Amin}}, \ and\ \bibinfo {author} {\bibfnamefont {A.~M.}\
  \bibnamefont {Zagoskin}},\ }\href {\doibase 10.1103/PhysRevLett.91.097906}
  {\bibfield  {journal} {\bibinfo  {journal} {Phys. Rev. Lett.}\ }\textbf
  {\bibinfo {volume} {91}},\ \bibinfo {pages} {097906} (\bibinfo {year}
  {2003})}\BibitemShut {NoStop}%
\bibitem [{\citenamefont {Hoi}\ \emph {et~al.}(2013)\citenamefont {Hoi},
  \citenamefont {Wilson}, \citenamefont {Johansson}, \citenamefont {Lindkvist},
  \citenamefont {Peropadre}, \citenamefont {Palomaki},\ and\ \citenamefont
  {Delsing}}]{Hoi_2013}%
  \BibitemOpen
  \bibfield  {author} {\bibinfo {author} {\bibfnamefont {I.-C.}\ \bibnamefont
  {Hoi}}, \bibinfo {author} {\bibfnamefont {C.~M.}\ \bibnamefont {Wilson}},
  \bibinfo {author} {\bibfnamefont {G.}~\bibnamefont {Johansson}}, \bibinfo
  {author} {\bibfnamefont {J.}~\bibnamefont {Lindkvist}}, \bibinfo {author}
  {\bibfnamefont {B.}~\bibnamefont {Peropadre}}, \bibinfo {author}
  {\bibfnamefont {T.}~\bibnamefont {Palomaki}}, \ and\ \bibinfo {author}
  {\bibfnamefont {P.}~\bibnamefont {Delsing}},\ }\href {\doibase
  10.1088/1367-2630/15/2/025011} {\bibfield  {journal} {\bibinfo  {journal}
  {New Journal of Physics}\ }\textbf {\bibinfo {volume} {15}},\ \bibinfo
  {pages} {025011} (\bibinfo {year} {2013})}\BibitemShut {NoStop}%
\bibitem [{\citenamefont {Sowa1}\ and\ \citenamefont {Zagoskin}()}]{artur}%
  \BibitemOpen
  \bibfield  {author} {\bibinfo {author} {\bibfnamefont {A.~P.}\ \bibnamefont
  {Sowa1}}\ and\ \bibinfo {author} {\bibfnamefont {A.~M.}\ \bibnamefont
  {Zagoskin}},\ }\href@noop {} {\bibinfo  {journal} {arXiv:1902.05324}\
  }\BibitemShut {NoStop}%
\bibitem [{\citenamefont {Rakhmanov}\ \emph {et~al.}(2008)\citenamefont
  {Rakhmanov}, \citenamefont {Zagoskin}, \citenamefont {Savel'ev},\ and\
  \citenamefont {Nori}}]{rakhmanov2008quantum}%
  \BibitemOpen
\bibfield  {journal} {  }\bibfield  {author} {\bibinfo {author} {\bibfnamefont
  {A.~L.}\ \bibnamefont {Rakhmanov}}, \bibinfo {author} {\bibfnamefont {A.~M.}\
  \bibnamefont {Zagoskin}}, \bibinfo {author} {\bibfnamefont {S.}~\bibnamefont
  {Savel'ev}}, \ and\ \bibinfo {author} {\bibfnamefont {F.}~\bibnamefont
  {Nori}},\ }\href@noop {} {\bibfield  {journal} {\bibinfo  {journal} {Physical
  Review B}\ }\textbf {\bibinfo {volume} {77}},\ \bibinfo {pages} {144507}
  (\bibinfo {year} {2008})}\BibitemShut {NoStop}%
\bibitem [{\citenamefont {Zagoskin}\ \emph {et~al.}(2009)\citenamefont
  {Zagoskin}, \citenamefont {Rakhmanov}, \citenamefont {Savel'ev},\ and\
  \citenamefont {Nori}}]{zagoskin2009quantum}%
  \BibitemOpen
  \bibfield  {author} {\bibinfo {author} {\bibfnamefont {A.}~\bibnamefont
  {Zagoskin}}, \bibinfo {author} {\bibfnamefont {A.}~\bibnamefont {Rakhmanov}},
  \bibinfo {author} {\bibfnamefont {S.}~\bibnamefont {Savel'ev}}, \ and\
  \bibinfo {author} {\bibfnamefont {F.}~\bibnamefont {Nori}},\ }\href@noop {}
  {\bibfield  {journal} {\bibinfo  {journal} {physica status solidi (b)}\
  }\textbf {\bibinfo {volume} {246}},\ \bibinfo {pages} {955} (\bibinfo {year}
  {2009})}\BibitemShut {NoStop}%
\bibitem [{\citenamefont {Savel'ev}\ \emph {et~al.}(2012)\citenamefont
  {Savel'ev}, \citenamefont {Zagoskin}, \citenamefont {Rakhmanov},
  \citenamefont {Omelyanchouk}, \citenamefont {Washington},\ and\ \citenamefont
  {Nori}}]{savel2012two}%
  \BibitemOpen
  \bibfield  {author} {\bibinfo {author} {\bibfnamefont {S.}~\bibnamefont
  {Savel'ev}}, \bibinfo {author} {\bibfnamefont {A.}~\bibnamefont {Zagoskin}},
  \bibinfo {author} {\bibfnamefont {A.}~\bibnamefont {Rakhmanov}}, \bibinfo
  {author} {\bibfnamefont {A.}~\bibnamefont {Omelyanchouk}}, \bibinfo {author}
  {\bibfnamefont {Z.}~\bibnamefont {Washington}}, \ and\ \bibinfo {author}
  {\bibfnamefont {F.}~\bibnamefont {Nori}},\ }\href@noop {} {\bibfield
  {journal} {\bibinfo  {journal} {Physical Review A}\ }\textbf {\bibinfo
  {volume} {85}},\ \bibinfo {pages} {013811} (\bibinfo {year}
  {2012})}\BibitemShut {NoStop}%
\bibitem [{\citenamefont {Bamba}\ \emph {et~al.}(2016)\citenamefont {Bamba},
  \citenamefont {Inomata},\ and\ \citenamefont
  {Nakamura}}]{bamba2016superradiant}%
  \BibitemOpen
  \bibfield  {author} {\bibinfo {author} {\bibfnamefont {M.}~\bibnamefont
  {Bamba}}, \bibinfo {author} {\bibfnamefont {K.}~\bibnamefont {Inomata}}, \
  and\ \bibinfo {author} {\bibfnamefont {Y.}~\bibnamefont {Nakamura}},\
  }\href@noop {} {\bibfield  {journal} {\bibinfo  {journal} {Physical review
  letters}\ }\textbf {\bibinfo {volume} {117}},\ \bibinfo {pages} {173601}
  (\bibinfo {year} {2016})}\BibitemShut {NoStop}%
\bibitem [{\citenamefont {Asai}\ \emph {et~al.}(2018)\citenamefont {Asai},
  \citenamefont {Kawabata}, \citenamefont {Savel?ev},\ and\ \citenamefont
  {Zagoskin}}]{asai2018quasi}%
  \BibitemOpen
  \bibfield  {author} {\bibinfo {author} {\bibfnamefont {H.}~\bibnamefont
  {Asai}}, \bibinfo {author} {\bibfnamefont {S.}~\bibnamefont {Kawabata}},
  \bibinfo {author} {\bibfnamefont {S.~E.}\ \bibnamefont {Savel?ev}}, \ and\
  \bibinfo {author} {\bibfnamefont {A.~M.}\ \bibnamefont {Zagoskin}},\
  }\href@noop {} {\bibfield  {journal} {\bibinfo  {journal} {The European
  Physical Journal B}\ }\textbf {\bibinfo {volume} {91}},\ \bibinfo {pages}
  {30} (\bibinfo {year} {2018})}\BibitemShut {NoStop}%
\bibitem [{\citenamefont {Ivi{\'c}}\ \emph {et~al.}(2016)\citenamefont
  {Ivi{\'c}}, \citenamefont {Lazarides},\ and\ \citenamefont
  {Tsironis}}]{ivic2016qubit}%
  \BibitemOpen
  \bibfield  {author} {\bibinfo {author} {\bibfnamefont {Z.}~\bibnamefont
  {Ivi{\'c}}}, \bibinfo {author} {\bibfnamefont {N.}~\bibnamefont {Lazarides}},
  \ and\ \bibinfo {author} {\bibfnamefont {G.}~\bibnamefont {Tsironis}},\
  }\href@noop {} {\bibfield  {journal} {\bibinfo  {journal} {Scientific
  reports}\ }\textbf {\bibinfo {volume} {6}},\ \bibinfo {pages} {29374}
  (\bibinfo {year} {2016})}\BibitemShut {NoStop}%
\bibitem [{\citenamefont {Lang}\ and\ \citenamefont
  {Sch\"utzhold}(2018)}]{lang2018}%
  \BibitemOpen
  \bibfield  {author} {\bibinfo {author} {\bibfnamefont {S.}~\bibnamefont
  {Lang}}\ and\ \bibinfo {author} {\bibfnamefont {R.}~\bibnamefont
  {Sch\"utzhold}},\ }\href@noop {} {\bibfield  {journal} {\bibinfo  {journal}
  {{\tt arXiv:1808.07453}}\ } (\bibinfo {year} {2018})}\BibitemShut {NoStop}%
\bibitem [{\citenamefont {Tian}\ \emph {et~al.}(2017)\citenamefont {Tian},
  \citenamefont {Jing},\ and\ \citenamefont {Dragan}}]{Tian_2017}%
  \BibitemOpen
  \bibfield  {author} {\bibinfo {author} {\bibfnamefont {Z.}~\bibnamefont
  {Tian}}, \bibinfo {author} {\bibfnamefont {J.}~\bibnamefont {Jing}}, \ and\
  \bibinfo {author} {\bibfnamefont {A.}~\bibnamefont {Dragan}},\ }\href
  {\doibase 10.1103/PhysRevD.95.125003} {\bibfield  {journal} {\bibinfo
  {journal} {Phys. Rev. D}\ }\textbf {\bibinfo {volume} {95}},\ \bibinfo
  {pages} {125003} (\bibinfo {year} {2017})}\BibitemShut {NoStop}%
\bibitem [{\citenamefont {Wilson}\ \emph {et~al.}(2011)\citenamefont {Wilson},
  \citenamefont {Johansson}, \citenamefont {Pourkabirian}, \citenamefont
  {Simoen}, \citenamefont {Johansson}, \citenamefont {Duty}, \citenamefont
  {Nori},\ and\ \citenamefont {Delsing}}]{wilson2011observation}%
  \BibitemOpen
  \bibfield  {author} {\bibinfo {author} {\bibfnamefont {C.}~\bibnamefont
  {Wilson}}, \bibinfo {author} {\bibfnamefont {G.}~\bibnamefont {Johansson}},
  \bibinfo {author} {\bibfnamefont {A.}~\bibnamefont {Pourkabirian}}, \bibinfo
  {author} {\bibfnamefont {M.}~\bibnamefont {Simoen}}, \bibinfo {author}
  {\bibfnamefont {J.}~\bibnamefont {Johansson}}, \bibinfo {author}
  {\bibfnamefont {T.}~\bibnamefont {Duty}}, \bibinfo {author} {\bibfnamefont
  {F.}~\bibnamefont {Nori}}, \ and\ \bibinfo {author} {\bibfnamefont
  {P.}~\bibnamefont {Delsing}},\ }\href@noop {} {\bibfield  {journal} {\bibinfo
   {journal} {Nature}\ }\textbf {\bibinfo {volume} {479}},\ \bibinfo {pages}
  {376} (\bibinfo {year} {2011})}\BibitemShut {NoStop}%
\bibitem [{\citenamefont {L\"ahteenm\"aki}\ \emph {et~al.}(2013)\citenamefont
  {L\"ahteenm\"aki}, \citenamefont {Paraoanu}, \citenamefont {Hassel},\ and\
  \citenamefont {Hakonen}}]{Laetheenmaeki_2013}%
  \BibitemOpen
  \bibfield  {author} {\bibinfo {author} {\bibfnamefont {P.}~\bibnamefont
  {L\"ahteenm\"aki}}, \bibinfo {author} {\bibfnamefont {G.~S.}\ \bibnamefont
  {Paraoanu}}, \bibinfo {author} {\bibfnamefont {J.}~\bibnamefont {Hassel}}, \
  and\ \bibinfo {author} {\bibfnamefont {P.}~\bibnamefont {Hakonen}},\
  }\bibfield  {booktitle} {\emph {\bibinfo {booktitle} {Proceedings of the
  National Academy of Sciences of the United States of America}},\ }\href@noop
  {} {\ \textbf {\bibinfo {volume} {110}},\ \bibinfo {pages} {4234} (\bibinfo
  {year} {2013})}\BibitemShut {NoStop}%
\bibitem [{\citenamefont {Grimsmo}\ and\ \citenamefont
  {Blais}(2017)}]{grimsmo2017squeezing}%
  \BibitemOpen
  \bibfield  {author} {\bibinfo {author} {\bibfnamefont {A.~L.}\ \bibnamefont
  {Grimsmo}}\ and\ \bibinfo {author} {\bibfnamefont {A.}~\bibnamefont
  {Blais}},\ }\href@noop {} {\bibfield  {journal} {\bibinfo  {journal} {npj
  Quantum Information}\ }\textbf {\bibinfo {volume} {3}},\ \bibinfo {pages}
  {20} (\bibinfo {year} {2017})}\BibitemShut {NoStop}%
\bibitem [{\citenamefont {Fang}\ and\ \citenamefont
  {Baranger}(2016)}]{FANG201692}%
  \BibitemOpen
  \bibfield  {author} {\bibinfo {author} {\bibfnamefont {Y.-L.~L.}\
  \bibnamefont {Fang}}\ and\ \bibinfo {author} {\bibfnamefont {H.~U.}\
  \bibnamefont {Baranger}},\ }\href {\doibase
  https://doi.org/10.1016/j.physe.2015.11.004} {\bibfield  {journal} {\bibinfo
  {journal} {Physica E: Low-dimensional Systems and Nanostructures}\ }\textbf
  {\bibinfo {volume} {78}},\ \bibinfo {pages} {92 } (\bibinfo {year}
  {2016})}\BibitemShut {NoStop}%
\bibitem [{\citenamefont {Fang}\ and\ \citenamefont
  {Baranger}(2017)}]{PhysRevA.96.013842}%
  \BibitemOpen
  \bibfield  {author} {\bibinfo {author} {\bibfnamefont {Y.-L.~L.}\
  \bibnamefont {Fang}}\ and\ \bibinfo {author} {\bibfnamefont {H.~U.}\
  \bibnamefont {Baranger}},\ }\href {\doibase 10.1103/PhysRevA.96.013842}
  {\bibfield  {journal} {\bibinfo  {journal} {Phys. Rev. A}\ }\textbf {\bibinfo
  {volume} {96}},\ \bibinfo {pages} {013842} (\bibinfo {year}
  {2017})}\BibitemShut {NoStop}%
\bibitem [{\citenamefont {Fang}\ and\ \citenamefont
  {Baranger}(2015)}]{PhysRevA.91.053845}%
  \BibitemOpen
  \bibfield  {author} {\bibinfo {author} {\bibfnamefont {Y.-L.~L.}\
  \bibnamefont {Fang}}\ and\ \bibinfo {author} {\bibfnamefont {H.~U.}\
  \bibnamefont {Baranger}},\ }\href {\doibase 10.1103/PhysRevA.91.053845}
  {\bibfield  {journal} {\bibinfo  {journal} {Phys. Rev. A}\ }\textbf {\bibinfo
  {volume} {91}},\ \bibinfo {pages} {053845} (\bibinfo {year}
  {2015})}\BibitemShut {NoStop}%
\bibitem [{\citenamefont {Lalumi\`ere}\ \emph {et~al.}(2013)\citenamefont
  {Lalumi\`ere}, \citenamefont {Sanders}, \citenamefont {van Loo},
  \citenamefont {Fedorov}, \citenamefont {Wallraff},\ and\ \citenamefont
  {Blais}}]{PhysRevA.88.043806}%
  \BibitemOpen
  \bibfield  {author} {\bibinfo {author} {\bibfnamefont {K.}~\bibnamefont
  {Lalumi\`ere}}, \bibinfo {author} {\bibfnamefont {B.~C.}\ \bibnamefont
  {Sanders}}, \bibinfo {author} {\bibfnamefont {A.~F.}\ \bibnamefont {van
  Loo}}, \bibinfo {author} {\bibfnamefont {A.}~\bibnamefont {Fedorov}},
  \bibinfo {author} {\bibfnamefont {A.}~\bibnamefont {Wallraff}}, \ and\
  \bibinfo {author} {\bibfnamefont {A.}~\bibnamefont {Blais}},\ }\href
  {\doibase 10.1103/PhysRevA.88.043806} {\bibfield  {journal} {\bibinfo
  {journal} {Phys. Rev. A}\ }\textbf {\bibinfo {volume} {88}},\ \bibinfo
  {pages} {043806} (\bibinfo {year} {2013})}\BibitemShut {NoStop}%
\bibitem [{\citenamefont {Zhang}\ \emph {et~al.}(2014)\citenamefont {Zhang},
  \citenamefont {Yu}, \citenamefont {Liang}, \citenamefont {Chen},
  \citenamefont {Jia},\ and\ \citenamefont {Nori}}]{zhang2014quantum}%
  \BibitemOpen
  \bibfield  {author} {\bibinfo {author} {\bibfnamefont {Y.}~\bibnamefont
  {Zhang}}, \bibinfo {author} {\bibfnamefont {L.}~\bibnamefont {Yu}}, \bibinfo
  {author} {\bibfnamefont {J.-Q.}\ \bibnamefont {Liang}}, \bibinfo {author}
  {\bibfnamefont {G.}~\bibnamefont {Chen}}, \bibinfo {author} {\bibfnamefont
  {S.}~\bibnamefont {Jia}}, \ and\ \bibinfo {author} {\bibfnamefont
  {F.}~\bibnamefont {Nori}},\ }\href@noop {} {\bibfield  {journal} {\bibinfo
  {journal} {Scientific reports}\ }\textbf {\bibinfo {volume} {4}},\ \bibinfo
  {pages} {4083} (\bibinfo {year} {2014})}\BibitemShut {NoStop}%
\bibitem [{\citenamefont {Mallet}\ \emph {et~al.}(2011)\citenamefont {Mallet},
  \citenamefont {Castellanos-Beltran}, \citenamefont {Ku}, \citenamefont
  {Glancy}, \citenamefont {Knill}, \citenamefont {Irwin}, \citenamefont
  {Hilton}, \citenamefont {Vale},\ and\ \citenamefont {Lehnert}}]{HD}%
  \BibitemOpen
  \bibfield  {author} {\bibinfo {author} {\bibfnamefont {F.}~\bibnamefont
  {Mallet}}, \bibinfo {author} {\bibfnamefont {M.~A.}\ \bibnamefont
  {Castellanos-Beltran}}, \bibinfo {author} {\bibfnamefont {H.~S.}\
  \bibnamefont {Ku}}, \bibinfo {author} {\bibfnamefont {S.}~\bibnamefont
  {Glancy}}, \bibinfo {author} {\bibfnamefont {E.}~\bibnamefont {Knill}},
  \bibinfo {author} {\bibfnamefont {K.~D.}\ \bibnamefont {Irwin}}, \bibinfo
  {author} {\bibfnamefont {G.~C.}\ \bibnamefont {Hilton}}, \bibinfo {author}
  {\bibfnamefont {L.~R.}\ \bibnamefont {Vale}}, \ and\ \bibinfo {author}
  {\bibfnamefont {K.~W.}\ \bibnamefont {Lehnert}},\ }\href {\doibase
  10.1103/PhysRevLett.106.220502} {\bibfield  {journal} {\bibinfo  {journal}
  {Phys. Rev. Lett.}\ }\textbf {\bibinfo {volume} {106}},\ \bibinfo {pages}
  {220502} (\bibinfo {year} {2011})}\BibitemShut {NoStop}%
\bibitem [{\citenamefont {Navez}\ \emph {et~al.}(2001)\citenamefont {Navez},
  \citenamefont {Brambilla}, \citenamefont {Gatti},\ and\ \citenamefont
  {Lugiato}}]{PhysRevA.65.013813}%
  \BibitemOpen
  \bibfield  {author} {\bibinfo {author} {\bibfnamefont {P.}~\bibnamefont
  {Navez}}, \bibinfo {author} {\bibfnamefont {E.}~\bibnamefont {Brambilla}},
  \bibinfo {author} {\bibfnamefont {A.}~\bibnamefont {Gatti}}, \ and\ \bibinfo
  {author} {\bibfnamefont {L.~A.}\ \bibnamefont {Lugiato}},\ }\href {\doibase
  10.1103/PhysRevA.65.013813} {\bibfield  {journal} {\bibinfo  {journal} {Phys.
  Rev. A}\ }\textbf {\bibinfo {volume} {65}},\ \bibinfo {pages} {013813}
  (\bibinfo {year} {2001})}\BibitemShut {NoStop}%
\bibitem [{\citenamefont {Einstein}\ \emph {et~al.}(1935)\citenamefont
  {Einstein}, \citenamefont {Podolsky},\ and\ \citenamefont
  {Rosen}}]{PhysRev.47.777}%
  \BibitemOpen
  \bibfield  {author} {\bibinfo {author} {\bibfnamefont {A.}~\bibnamefont
  {Einstein}}, \bibinfo {author} {\bibfnamefont {B.}~\bibnamefont {Podolsky}},
  \ and\ \bibinfo {author} {\bibfnamefont {N.}~\bibnamefont {Rosen}},\ }\href
  {\doibase 10.1103/PhysRev.47.777} {\bibfield  {journal} {\bibinfo  {journal}
  {Phys. Rev.}\ }\textbf {\bibinfo {volume} {47}},\ \bibinfo {pages} {777}
  (\bibinfo {year} {1935})}\BibitemShut {NoStop}%
\bibitem [{\citenamefont {Zagoskin}(2011)}]{zagoskin2011quantum}%
  \BibitemOpen
  \bibfield  {author} {\bibinfo {author} {\bibfnamefont {A.~M.}\ \bibnamefont
  {Zagoskin}},\ }\href@noop {} {\emph {\bibinfo {title} {Quantum engineering:
  theory and design of quantum coherent structures}}}\ (\bibinfo  {publisher}
  {Cambridge University Press},\ \bibinfo {year} {2011})\BibitemShut {NoStop}%
\bibitem [{\citenamefont {Vool}\ and\ \citenamefont
  {Devoret}(2017)}]{vool2017introduction}%
  \BibitemOpen
  \bibfield  {author} {\bibinfo {author} {\bibfnamefont {U.}~\bibnamefont
  {Vool}}\ and\ \bibinfo {author} {\bibfnamefont {M.}~\bibnamefont {Devoret}},\
  }\href@noop {} {\bibfield  {journal} {\bibinfo  {journal} {International
  Journal of Circuit Theory and Applications}\ }\textbf {\bibinfo {volume}
  {45}},\ \bibinfo {pages} {897} (\bibinfo {year} {2017})}\BibitemShut
  {NoStop}%
\bibitem [{Note1()}]{Note1}%
  \BibitemOpen
  \bibinfo {note} {See the Supplemental Materials for the mathematical
  details}\BibitemShut {NoStop}%
\end{thebibliography}%


\end{document}